\documentclass[traditabstract]{aa}
\usepackage{graphicx}
\usepackage{natbib}
\usepackage{txfonts}
\begin{document}

   \title{Effective temperatures of red giants in the APOKASC catalogue and the mixing length calibration in stellar models}


   \author{M.~Salaris \inst{1}
          \and  S.~Cassisi \inst{2}
          \and R.~P.~Schiavon \inst{1} 
          \and A.~Pietrinferni \inst{2}
          }

   \institute{Astrophysics Research Institute,
     Liverpool John Moores University, IC2, Liverpool Science Park,
     146 Brownlow Hill, Liverpool, L3 5RF, UK (M.Salaris@ljmu.ac.uk) 
     \and INAF-Osservatorio Astronomico d'Abruzzo, via M. Maggini, sn.
     64100, Teramo, Italy}

   \date{Received ; accepted }

 
\abstract{Red giants in the updated APOGEE-$Kepler$ catalogue, with estimates of mass, chemical composition, surface gravity and effective temperature, 
have recently challenged stellar models computed under the standard assumption of solar calibrated mixing length.      
In this work, we critically reanalyse this sample of red giants, adopting our own stellar model calculations. 
Contrary to previous results, we find that the disagreement 
between the $T_{\mathrm{eff}}$ scale of red giants and models with solar calibrated mixing length disappears when considering our models and 
the APOGEE-$Kepler$ stars with scaled solar metal distribution. 
However, a discrepancy shows up when $\alpha$-enhanced stars are included in the sample. 
We have found that assuming mass, chemical composition and effective temperature scale of the APOGEE-$Kepler$ catalogue, 
stellar models generally underpredict the change of temperature of red giants caused by $\alpha$-element enhancements at fixed [Fe/H].
A second important conclusion is that the choice of the outer boundary conditions employed in model calculations is critical. 
Effective temperature differences (metallicity dependent) between models with solar calibrated mixing length 
and observations appear for some choices of the boundary conditions, but 
this is not a general result.}
\titlerunning{APOKASC red giants and mixing length calibration}
\authorrunning{M. Salaris et al.}
   \maketitle
%

\section{Introduction}\label{intro}

Calculations of the superadiabatic convective temperature gradients 
in stellar evolution models are almost universally based on the very simple, local formalism provided by the mixing length theory \citep[MLT --][]{bv58}.
The convective flow is idealized in terms of columns of upwards and downwards moving elements all with the same characteristic size, 
that cover a fixed mean free path before dissolving. 
Both the mean free path and the characteristic size of the convective elements are assumed to be 
equal to $\Lambda$ = $\alpha_{\rm MLT} H_P$, the mixing length. The free parameter $\alpha_{\rm MLT}$ is typically 
assumed to be a constant value within the convective regions and along all evolutionary phases, 
and $H_p$ is the local pressure scale height. The chosen value of  $\alpha_{\rm MLT}$ determines the model $T_{\mathrm{eff}}$. 

It is well known that this simplistic MLT picture of convection is very different from 
results of two-dimensional (2D) and three-dimensional (3D) radiation hydrodynamics simulations of 
convection in stellar envelopes and atmospheres \citep[see, e.g.][and references therein]{sn, lfs, trampedach14, mwa15}. These computations 
show how convection consists mainly of continuous flows, with the warm gas rising almost adiabatically,
in a background of cool, narrower and faster downdrafts. 
A fraction of the upflows is continuously overturning to conserve mass on the background of the steep density gradients.

Clearly, we cannot expect the MLT to provide an accurate description of the thermal stratification within the 
superadiabatic layers of convective envelopes, but only an effective stratification that leads hopefully to an appropriate 
effective temperature ($T_{\mathrm{eff}}$) scale for the stellar models, once a suitable value of $\alpha_{\rm MLT}$ 
is chosen. This free parameter is 
usually calibrated by reproducing the radius of the Sun at the solar age with an evolutionary solar model \citep{gw76}. Of course 
there is no reason a priori why $\alpha_{\rm MLT}$ should be the same with varying mass, evolutionary phase, and chemical composition.

Additional free parameters appear in the MLT formalism, but they are generally fixed beforehand, giving origin to different flavours  
of the MLT formalism \citep[see, e.g.][and references therein]{pvi90, sc08}. 
Remarkably, different MLT flavours found in the literature provide essentially the same evolutionary tracks 
when $\alpha_{\rm MLT}$ is accordingly recalibrated on the Sun \citep{pvi90, sc08}.

One independent empirical way to calibrate and/or test whether the solar calibration of $\alpha_{\rm MLT}$ is appropriate 
also for other evolutionary phases/chemical compositions, is   
to compare empirically determined effective temperatures of red giant branch (RGB) stars, 
with theoretical models of the appropriate chemical composition, that are indeed very sensitive to the treatment of the superadiabatic layers  
\citep[see,.e.g.][and references therein]{sc:91,sc:96,vbs96}.

A very recent study by \citet{tayar} has analysed a sample of over 3000 RGB stars with $T_{\mathrm{eff}}$, mass, surface gravity ($g$), 
[Fe/H] and 
[$\alpha$/Fe] determinations from the updated APOGEE-{\sl Kepler} catalogue (APOKASC), suited for testing the mixing length calibration in 
theoretical stellar models. According to the grid of stellar evolution models specifically calculated to match the 
$T_{\mathrm{eff}}$ values of the individual stars, this study (hereafter T17) concluded that a variation of 
$\alpha_{\rm MLT}$ with varying [Fe/H] is required. Their stellar models with solar calibrated $\alpha_{\rm MLT}$ 
\citep[the solar $\alpha_{\rm MLT}$ in their calculations is equal to 1.72 when employing the][MLT flavour]{bv58} are unable to match the empirical  
$T_{\mathrm{eff}}$ values for [Fe/H] between $\sim+$0.4 and $-$1.0. They calculated 
differences $\Delta T\equiv T_{obs}-T_{models}$ between observed and theoretical $T_{\mathrm{eff}}$ for each individual star in their sample, 
and found that $\Delta T$=93.1[Fe/H]+ 107.5~K. They concluded that a varying mixing length 
$\alpha_{\rm MLT}=0.161 [Fe/H]+1.90$ is required to match the 
empirical temperatures. This relationship predicts a non solar $\alpha_{\rm MLT}$ also for RGB stars at solar metallicity.
The same authors found a similar trend of  $\Delta T$ with [Fe/H] (a $\Delta T$-[Fe/H] slope of about 100~K/dex) 
when using the PARSEC stellar models \citep[][]{parsec}, albeit with 
a zero point offset of about $-$100~K compared to the results obtained with their models.

A variation of $\alpha_{\rm MLT}$ with [Fe/H] --and potentially with evolutionary phase-- 
has obviously profound implications for the calibration of convection in stellar models, age estimates of RGB stars in 
the Hertzsprung-Russell and $g$-$T_{\mathrm{eff}}$ diagrams, and also stellar population integrated spectral features  
sensitive to the presence of a RGB component.

In light of the relevant implications of T17 result, we have reanalysed their APOKASC sample with our own independent stellar evolution calculations, 
paying particular attention to the role played by uncertanties in the calculation of the model boundary conditions. 
Our new results clarify the role played by the combination of $\alpha_{\rm MLT}$ and boundary conditions in the interpretations of the data, 
and, very importantly, discloses also a major difficulty when comparing models with T17 $\alpha$-enhanced stars.

Section~\ref{data} briefly summarizes T17 data and the models calculated for this work, followed 
in Sect.~\ref{analysis} by a description of our analysis and our new results. 
A summary and in-depth discussion of our findings closes the paper.

\section{Models and data}\label{data}

For our analysis we have calculated a large set of models using code and physics inputs employed 
to create the BaSTI database of stellar evolution models \citep[see][]{basti}. 
Because of the relevance to this work, we specify that radiative 
opacities are from the OPAL calculations \citep{ir:96} for temperatures larger than ${\rm \log(T)=4.0}$, 
whereas calculations by \cite{ferguson} --that include the  
contributions from molecules and grains-- have been adopted for lower temperatures. Both high- and low-temperature opacity tables 
account properly for the metal distributions adopted in our models (see below and Sect.~\ref{metdistr}).

We have just changed the $T(\tau)$ relation adopted in BaSTI to 
determine the models' outer boundary conditions (a crucial input for the determination of the models' $T_{\mathrm{eff}}$, as discussed in Sect.~\ref{bcon}),  
employing the \citet{vernazza:81} solar semi-empirical $T(\tau)$ (hereafter VAL) instead of the \citet{ks} one\footnote{We implement 
the following fit to \citet{vernazza:81} tabulation: $T^4 =0.75 \ T_{\mathrm{eff}}^4 \ (\tau+1.017-0.3 e^{-2.54 \tau}-
0.291 e^{-30 \tau})$, where $\tau$ is the Rosseland optical depth}. 
According to the analysis 
by \citet{sc15}, model tracks computed with this $T(\tau)$ relation approximate well results obtained using the hydro-calibrated $T(\tau)$
relationships provided by \citet{trampedach14} for the solar chemical composition.
We will come back to this issue in Sect.~\ref{bcon}.

We have computed a model grid for masses between 0.7 and 2.6~$M_{\odot}$ in 0.1~$M_{\odot}$ increments, and 
scaled solar [Fe/H] between $-$2.0 and $+0.4$~dex in steps of 0.2-0.3~dex. A solar model including atomic diffusion 
has been calibrated to determine initial solar values of He and metal mass fractions 
$Y$=0.274, $Z$=0.0199 \citep[for the][solar metal mixture]{gn93}, and mixing length \citep[with the MLT flavour from][]{bv58} $\alpha_{\rm MLT}$=1.90. 
We have adopted the same $Y$-$Z$ relationship $Y$=0.245+1.41 $\times Z$ as in BaSTI. Our model grid does not include atomic diffusion, because 
its effect on the RGB evolution is, as well known, negligible \citep{sc15}.
In addition, we have calculated sets of $\alpha$-enhanced models for the various masses and [Fe/H] of the grid, and [$\alpha$/Fe]=0.4.

With these models we have first reanalysed the sample of RGB field stars by T17, considering the  
log($g$), $T_{\mathrm{eff}}$, [Fe/H] and [$\alpha$/Fe] values listed in the publicly available data file.
Masses are derived from asteroseismic scaling relations, and the other quantities are obtained 
using the APOGEE spectroscopic data set. The $T_{\mathrm{eff}}$ values are calibrated to be consistent with the 
\citet{ghb09} effective temperature scale, based on the infrared-flux method.

We considered only the stars with a calculated error bar on the mass determinations 
(we excluded objects with error on the mass given as -9999), that still leave a sample of well over 3000 objects, 
spanning a mass range between 0.8 and 2.4 $M_{\odot}$, with a strong peak of the mass distribution around 1.2-1.3$M_{\odot}$.

   \begin{figure}
   \centering
   \includegraphics[width=8.7cm]{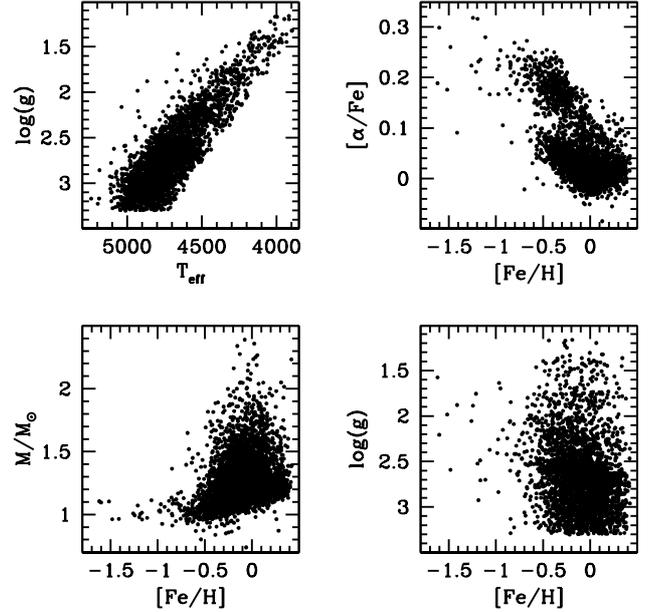}
      \caption{The adopted sample of RGB stars from the APOKASC dataset displayed in various diagrams: $\log(g) - T_{\mathrm{eff}}$,
      ${\rm [\alpha/Fe] - [Fe/H]}$, ${\rm mass - [Fe/H]}$ and $\log(g) - {\rm [Fe/H]}$.}
         \label{sample}
   \end{figure}
  
Figure~\ref{sample} displays the data in four different diagrams. The stars 
cover a log($g$) range between $\sim$3.3 and 1.1 (in cgs units), and $T_{\mathrm{eff}}$ between $\sim$5200 and 3900~K, with 
the bulk of the stars having [Fe/H] between $\sim -$0.7 and $\sim$ +0.4~dex, and a maximum $\alpha$-enhancement 
typically around 0.25~dex.
Notice that stars with a given [Fe/H] typically cover the full range of surface gravities, 
but the range of masses at a given [Fe/H] varies with metallicity, due to the variation of the age distribution 
of Galactic disk stars with [Fe/H]. 

To determine differences $\Delta T\equiv T_{obs}-T_{models}$ between observed and theoretical $T_{\mathrm{eff}}$ for each individual star in T17 sample, 
we have interpolated linearly in mass, [Fe/H], [$\alpha$/Fe], log($g$) amongst the models, 
to determine the corresponding theoretical $T_{\mathrm{eff}}$ for each observed star. 

\section{Analysis of T17 sample}\label{analysis}

\begin{figure}
   \centering
   \includegraphics[width=8.7cm]{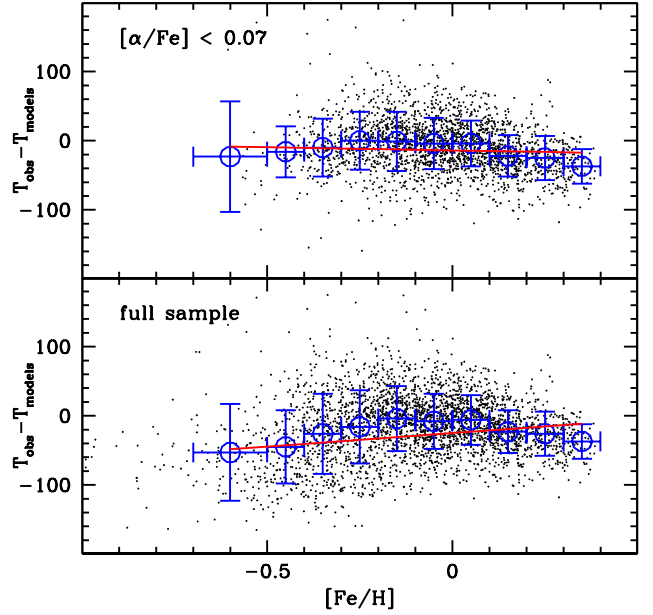}
      \caption{$\Delta T$ as a function of [Fe/H] (dots) for the whole sample of RGB stars (bottom panel) and for a sub-sample of RGB objects with 
      ${\rm [\alpha/Fe]<0.07}$ (top panel). Open circles with error bars denote the mean values of $\Delta T$  
      in specific metallicity bins --Columns 3 and 4 in Tables~\ref{tab01} and ~\ref{tab02}-- 
      while the solid lines display linear fits to the binned data.}
         \label{differences}
   \end{figure}

The top panel of Fig.~\ref{differences} displays the differences $\Delta T\equiv T_{obs}-T_{models}$ as a function of [Fe/H] for the full T17 sample; 
we have considered stars with [Fe/H]$> -$0.7, to include only the [Fe/H] range well sampled by the data. 
It is easy to notice a trend of $\Delta T$ with [Fe/H], qualitatively similar to what found by T17. 
We have overplotted, as open circles, mean values of $\Delta T$ determined 
in ten [Fe/H] bins, with total width of 0.10~dex, apart from the most metal poor bin, that has a 
width of 0.20~dex, due to the smaller number of stars populating that metallicity range --see Table~\ref{tab01}. 
The horizontal error bars cover the width of the individual 
bins, while the vertical ones denote the 1$\sigma$ dispersion of $\Delta T$ around the mean values. 

\begin{table}
  \caption{Mean values of the differences $\Delta T$ (third column) for the full T17 sample, determined in [Fe/H] bins centred 
around the values given in the first column of the table, and half-widths listed in the second column. 
The last column displays the 1$\sigma$ dispersion around the mean  $\Delta T$ values.
\label{tab01}}
  \begin{tabular}{c c c c}
[Fe/H] &  $\pm \Delta[Fe/H]$  & $<\Delta T>$ (K) & $\sigma (\Delta T$) (K) \\
\hline
  0.35  &0.05    &  -35  & 24 \\
  0.25  &0.05   &   -23  & 33 \\
  0.15  &0.05   &   -21  & 31 \\ 
  0.05  &0.05  &     -8  & 35 \\
 -0.05  &0.05   &   -11  & 40 \\
 -0.15  &0.05   &    -9  & 44 \\
 -0.25  &0.05  &    -22  & 52 \\
 -0.35  &0.05  &    -32  & 54 \\         
 -0.45  &0.05   &   -54  & 53 \\ 
 -0.60  &0.10   &   -72  & 55 \\ 
\hline
\end{tabular}
\end{table}

As in T17, we find a drop of $\Delta T$ with decreasing [Fe/H], when [Fe/H] is below $\sim-$0.25. If we perform a simple 
linear fit through these mean values (considering the 1$\sigma$ dispersion as the error on these mean $\Delta T$ values) 
we obtain for the full sample 
$\Delta T$=(39 $\pm$ 19) [Fe/H]$-$(25 $\pm$ 6)~K, valid over a [Fe/H] range of $\sim$1.1~dex. 
A linear fit is clearly not the best approximation of the $\Delta T$-[Fe/H] global trend --as mentioned also in T17-- but 
it replicates T17 analysis and suffices to highlight 
the main results of these comparisons.

\begin{figure}
   \centering
   \includegraphics[width=8.7cm]{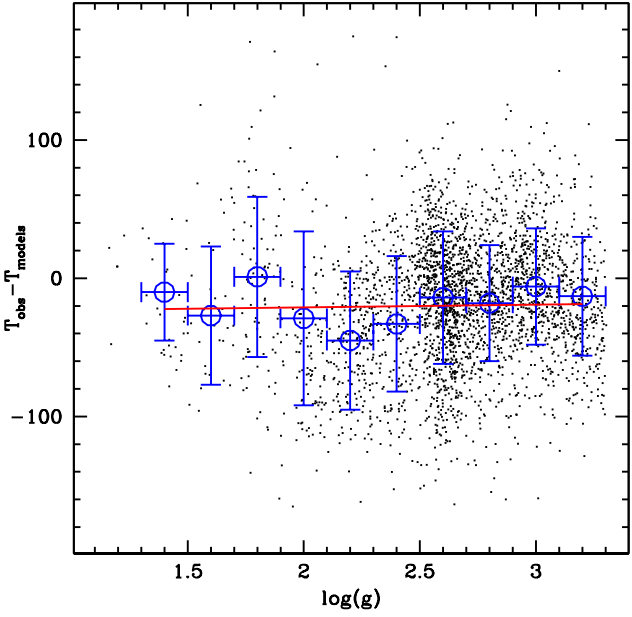}
      \caption{$\Delta T$ as a function of log($g$) for the whole sample of RGB stars. Open circles with error bars denote the mean values of $\Delta T$  
      in 0.2~dex log($g$) bins, and the 1$\sigma$ dispersion around these mean values. The solid line displays a linear fit to the binned data, with a slope that is 
      consistent with zero (the slope is equal to 2 $\pm$ 8 K/dex).}
         \label{differencesg}
   \end{figure}

Our RGB models turn out to be systematically hotter than observations by just 25~K at solar [Fe/H], 
a negligible value considering the error on the \citet{ghb09} $T_{\mathrm{eff}}$ calibration  
(the quoted average error on their RGB $T_{\mathrm{eff}}$ scale is $\leq$76~K), but the differences increase 
with decreasing [Fe/H].  
The slope we derive is about half the value of the slope determined by T17 with their own calculations, and the zero point is 
about 130~K lower. We do not find any correlation between $\Delta T$ and the surface gravity $g$ of the observed stars, as shown 
in Fig.~\ref{differencesg}.

\begin{table}
  \caption{As Table~\ref{tab01}, but for stars with observed [$\alpha$/Fe]$<$0.07.\label{tab02}}
  \begin{tabular}{c c c c}
[Fe/H] &  $\pm \Delta[Fe/H]$  & $<\Delta T>$ (K) & $\sigma (\Delta T)$ (K) \\
\hline
  0.35  &0.05    &  -34  & 23 \\
  0.25  &0.05   &   -22  & 30\\
  0.15  &0.05    &  -18  & 29\\
  0.05  &0.05   &    -4  & 32\\
 -0.05  &0.05    &   -4 &  37\\
 -0.15  &0.05   &     0  & 39\\
 -0.25  &0.05   &     0  & 39\\
 -0.35  &0.05   &    -7  & 38 \\       
 -0.45  &0.05  &     -7  & 43 \\
 -0.60  &0.10   &   -34  & 52 \\
\hline
\end{tabular}
\end{table}

The lower panel of Fig.~\ref{differences} displays $\Delta T$ values considering this time only stars with 
essentially scaled solar metal mixture, that is [$\alpha$/Fe]$<$0.07 (we chose this upper limit 
that is approximately equal to five times the 1$\sigma$ error on [$\alpha$/Fe] quoted in T17 data, but an upper limit closer to zero 
does not change the results). The overplotted mean values in the various [Fe/H] bins are reported in Table~\ref{tab02}.  
This time the trend of $\Delta T$ with [Fe/H] is not statistically significant. A linear fit to the mean $\Delta T$ values provides 
$\Delta T$=($-$9 $\pm$ 15) [Fe/H]$-$(14 $\pm$ 5), meaning that now theory and observations are essentially in agreement.

\begin{figure}
   \centering
   \includegraphics[width=8.7cm]{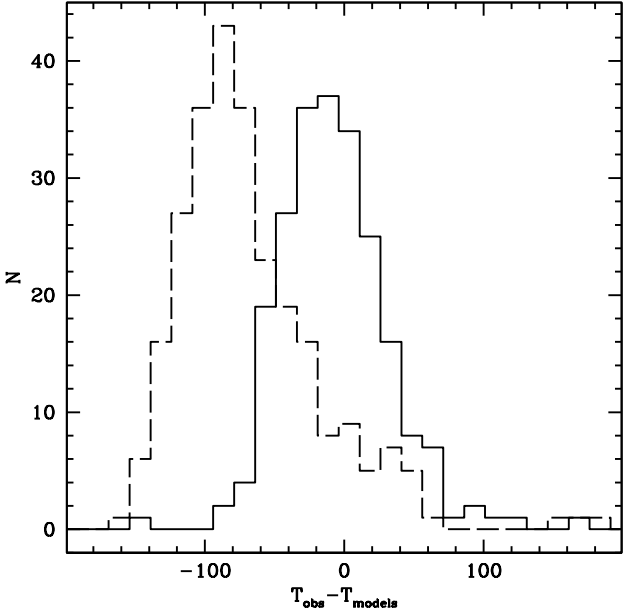}
      \caption{Histograms $\Delta T$ for stars with [Fe/H] between $-$0.5 and $-$0.3, [$\alpha$/Fe]$<0.07$ (solid line) and [$\alpha$/Fe]$\ge$0.07 (dashed line)}
         \label{differencesafe}
   \end{figure}

Figure~\ref{differencesafe} makes clearer the reason for the different result obtained when neglecting the $\alpha$-enhanced stars.  
We show here histograms of $\Delta T$ values for stars with [Fe/H] between $-$0.5 and $-$0.3, 
[$\alpha$/Fe]$<$0.07 and [$\alpha$/Fe]$\ge$0.07, respectively. One can clearly see how, in the same [Fe/H] range, 
we determine for $\alpha$-enhanced stars systematically lower $\Delta T$ values.

In conclusion, with our calculations the trend of $\Delta T$ with [Fe/H] is introduced by the inability of $\alpha$-enhanced models with solar calibrated 
$\alpha_{\rm MLT}$ to match 
their observational counteparts, that is stellar models are increasingly hotter than observations when [$\alpha$/Fe] 
increases, at fixed [Fe/H], even 
though we have taken into account the theoretically expected effect of [$\alpha$/Fe] on the model $T_{\mathrm{eff}}$, at a given [Fe/H].
Comparing the values in Tables~\ref{tab01} and \ref{tab02} one can notice that the effect of excluding $\alpha$-enhanced stars on 
the mean $\Delta T$ values appears around [Fe/H]=$-$0.25, consistent with the fact that below this [Fe/H] the fraction of  
$\alpha$-enhanced stars increases, and [$\alpha$/Fe] values also increase. 
On the other hand, and very importantly, our models with solar calibrated $\alpha_{\rm MLT}$ produce $T_{\mathrm{eff}}$ values for scaled solar stars that are 
generally consistent with observations over a [Fe/H] range of about 1~dex.

\begin{figure}
   \centering
   \includegraphics[width=8.7cm]{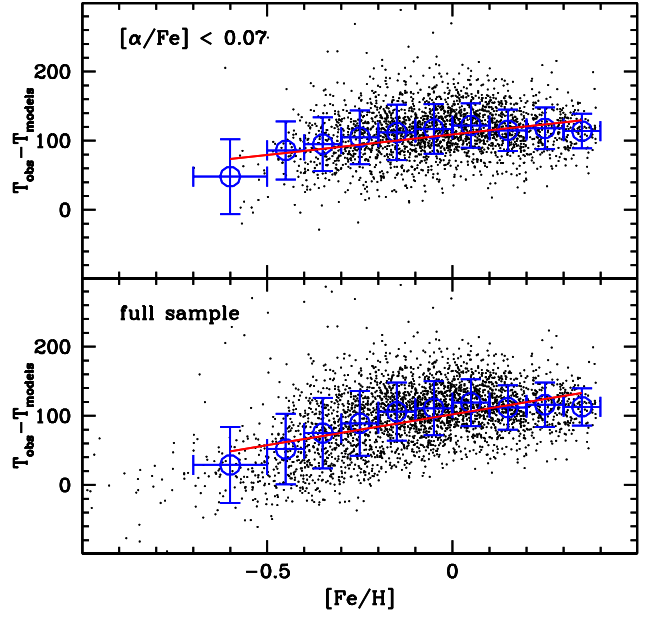}
      \caption{As Fig.~\ref{differences} but in this case individual $\Delta T$ values  
       are from the study by T17.}
         \label{dTTayar}
   \end{figure}

Our conclusions appear to be very different from T17 results obtained with their own model calculations, and therefore 
we have reanalysed T17 temperature differences (with respect ot their own solar calibrated $\alpha_{\rm MLT}$  models) 
on a star-by-star basis as provided by the authors\footnote{T17 did not provide 
the $\Delta T$ values they obtained employing the PARSEC models}, using exactly the same 
[Fe/H] bins discussed before. The results are displayed in Fig.~\ref{dTTayar}. 
T17 $\Delta T$ values have been binned in the same [Fe/H] ranges employed for our own results, and 
we have then performed a linear fit to 
the mean values of the ten [Fe/H] bins for both the full sample, and the sample restricted to stars with  
[$\alpha$/Fe]$<$0.07.

The linear fit to the full  sample provides 
$\Delta T$=(89 $\pm$ 16) [Fe/H]+(102 $\pm$ 5)~K. Slope and zero point are well consistent 
with the values (93.1~K/dex for the slope and 107.5~K for the zero point) determined by T17 
using their  different --finer-- binning of the data. This means that the slight different way to 
analyse $\Delta T$ values as employed in our analysis, provides exactly the same results found by T17, 
when applied to T17 estimates of $\Delta T$. 

Restricting the sample to objects with [$\alpha$/Fe]$<$0.07, the linear fit to T17 $\Delta T$ values 
provides $\Delta T$=(59 $\pm$ 14) [Fe/H]+(109 $\pm$ 5)~K. 
A slope different from zero is still present, contrary to what we find with our calculations, hence it 
cannot be attributed to just an inconsistent modelling of the $\alpha$-enhanced population. 
On the other hand, this slope is lower than the case of the full sample, and implies that the match 
of $\alpha$-enhanced stars with solar $\alpha_{\rm MLT}$ models increases the trend of $\Delta T$ with [Fe/H],   
compared to the case of just stars with scaled solar metal composition.

\begin{figure}
   \centering
   \includegraphics[width=8.7cm]{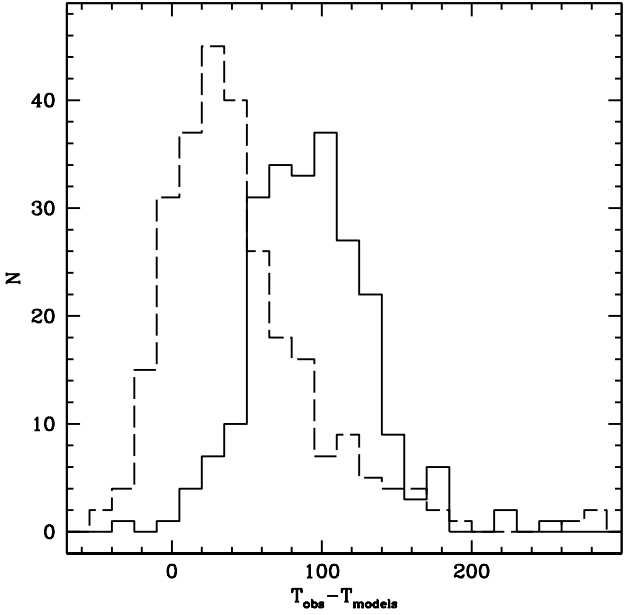}
      \caption{As Fig.~\ref{differencesafe} but for the $\Delta T$ values from T17}
         \label{differencesafeT}
   \end{figure}

This is exemplified by Fig.~\ref{differencesafeT}, that is the same as Fig.~\ref{differencesafe}, this time considering T17 $\Delta T$ values.
One can see clearly that also in case of T17 models, $\alpha$-enhanced stars at the same [Fe/H] display different (lower) 
$\Delta T$ compared to the scaled solar counterparts, exactly as in case of our models. 

It is also important to notice also a large difference, of about 120~K, in the zero points compared to our results.

\subsection{Revisiting the chemical composition of T17 stars}\label{analysis_2}

In light of the inconsistency between our solar calibrated $\alpha_{\rm MLT}$ models for $\alpha$-enhanced compositions  
and the observed $\alpha$-enhanced stars, 
we have investigated in more detail the chemical composition of T17 sample, looking at the abundances reported in 
the APOGEE DR13 catalogue \citep[][and Holtzman et al., in preparation]{maj}.

We have realized that [Fe/H] values reported by T17 are actually labelled as [M/H] in the DR13 catalogue, 
and [$\alpha$/Fe] is actually labelled as [$\alpha$/M] 
in DR13. As explained in the APOGEE catalogue\footnote{http://www.sdss.org/dr12/irspec/aspcap/}, 
the listed [M/H] is an overall scaling of 
metal abundances assuming a solar abundance ratio pattern, and [$\alpha$/M]=  [$\alpha$/H] -  ${\rm [M/H]}_{\odot}$.  
This definition of [M/H] cannot be implemented in stellar evolution calculations in a straightforward fashion 
for $\alpha$-enhanced metal mixtures, therefore 
we have extracted from the DR13 catalogue values for [Fe/H], and the listed $\alpha$-element abundance ratios [O/Fe], [Mg/Fe], 
[Si/Fe], [Ca/Fe] and [Ti/Fe] for all stars in T17 sample.

\begin{figure}
   \centering
   \includegraphics[width=8.7cm]{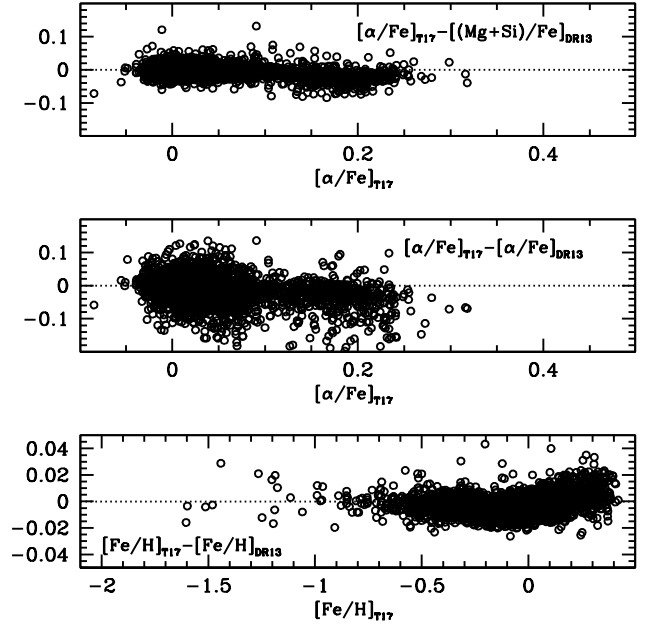}
      \caption{{\sl Bottom panel}: Difference between [Fe/H] values adopted by T17 (${\rm [Fe/H]_{T17}}$) 
      and [Fe/H] as listed in the DR13 catalogue (${\rm [Fe/H]_{D13}}$), as a function of ${\rm [Fe/H]_{T17}}$. 
      {\sl Middle panel}: Same as bottom panel, but for ${\rm [\alpha/Fe]}$, considering the total $\alpha$-enhancement from 
      DR13 data (see text for details).
      {\sl Top panel}: Same as middle panel, but in this case the DR13 estimate is actually [(Mg+Si)/Fe] (see text for details).}
         \label{compchim}
   \end{figure}

The bottom panel of Fig.~\ref{compchim} compares these [Fe/H] values with the values listed 
by T17 (that correspond to [M/H] in DR13). The agreement is typically within $\pm$0.02~dex, and 
suprisingly also for the $\alpha$-enhanced stars. At any rate, the consequence is that 
the general agreement of the $T_{\mathrm{eff}}$ of our solar $\alpha_{\rm MLT}$ RGB models with the observed scaled solar metallicy stars is confirmed (as we have 
verified applying the preocedure described in the previous section, employing these  DR13 [Fe/H] values) 
when using the DR13 values labelled as [Fe/H]. 

The top and middle panels of Fig.~\ref{compchim} compare the [$\alpha$/Fe] values given by T17 (that correspond to [$\alpha$/M] in DR13 catalogue) 
with two different estimates of [$\alpha$/Fe] based upon the DR13 values of [O/Fe], [Mg/Fe], 
[Si/Fe], [Ca/Fe] and [Ti/Fe]. 
In the top panel we display our [$\alpha$/Fe] values estimated as [(Mg+Si)/Fe], taking into account that 
Mg and Si are the two $\alpha$-elements that affect the $T_{\mathrm{eff}}$ of RGB stellar evolution models \citep{vdb12}. 
We have calculated [(Mg+Si)/Fe] employing the observed [Mg/Fe] and [Si/Fe], and the 
\cite{gn93} solar metal mixture used in the model calculations as a reference.
The correspondence with [$\alpha$/Fe] values listed by T17 (actually [$\alpha$/M] in DR13) is remarkable, 
with an offset of typically just $\sim$0.01~dex when [$\alpha$/Fe]$>$0.1, and a very small spread.

The middle panel displays our [$\alpha$/Fe] estimates accounting for all DR13 $\alpha$-elements. 
On average our [$\alpha$/Fe] tend to get systematically larger than T17 values , when [$\alpha$/Fe] is larger than $\sim$0.1~dex, 
again by just 0.01-0.02~dex on average. However, there is now a large scatter (compared to the case of [(Mg+Si)/Fe]) in the differences.    
The reason 
for the different behaviour compared to the [(Mg+Si)/Fe] case is 
that the various $\alpha$ elements are not always enhanced by the same amount in the observed stars, hence the exact value of [$\alpha$/Fe] 
to some degree depends on which elements are included in its definition.

From the point of view of testing the RGB model $T_{\mathrm{eff}}$, what matters is that the chemical composition of the 
models match the observed [(Mg+Si)/Fe] \citep[based on the results by][]{vdb12}. In case of the same enhancement for 
all $\alpha$ elements, typical of stellar evolution calculations, 
this means that the model [$\alpha$/Fe] has to match the observed [(Mg+Si)/Fe]. Therefore the  
results of the comparison made in the previous section employing T17 [$\alpha$/Fe] values still stand, given that T17 [$\alpha$/Fe] 
corresponds very closely to [(Mg+Si)/Fe] as determined from the DR13 individual abundances.

\begin{figure}
   \centering
   \includegraphics[width=8.7cm]{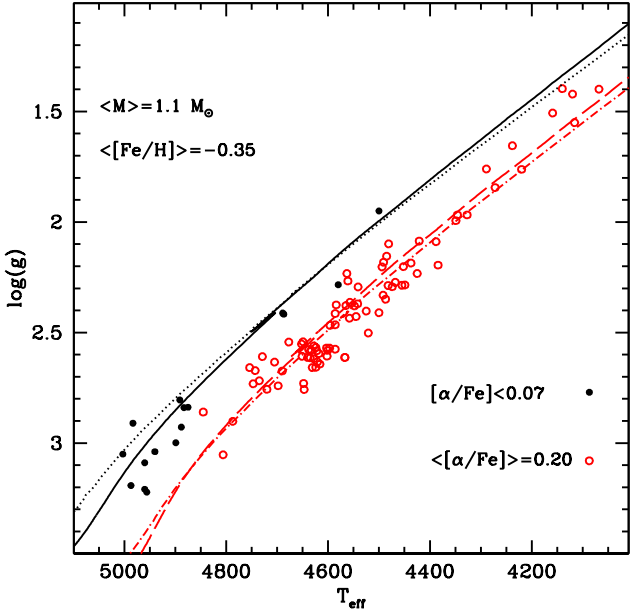}
      \caption{${\log(g) - T_{\mathrm{eff}}}$ diagram of two sub-samples of APOKASC RGB stars with average ${\rm [Fe/H]=-0.35}$ and mass ${\rm 1.1M_\odot}$, 
but different $\alpha-$enhancements: scaled solar objects ([$\alpha$/Fe]$<$0.07,  filled circles) 
 and $\alpha-$enhanced  (average ${\rm [\alpha/Fe]=0.2}$, open circles). Our ${\rm 1.1M_\odot}$, [Fe/H]=$-$0.35 evolutionary tracks 
      for ${\rm [\alpha/Fe]=0}$ (solid line) and $+0.40$ (dashed line) are also displayed. The $\alpha$-enhancement of the theoretical models 
      is twice the average enhancement of the selected $\alpha$-enhanced stars. 
      Dotted and dot-dashed lines show tracks with the same mass, [Fe/H] and $\alpha$-enhancements, from \cite{dartmouth}.}
         \label{dtaenfe}
   \end{figure}

Finally, Fig.~\ref{dtaenfe} show very clearly the problem when matching the $T_{\mathrm{eff}}$ of $\alpha$-enhanced stars with models. 
We have displayed the log($g$)-$T_{\mathrm{eff}}$ diagram of two samples of stars with observed 
mean mass equal to 1.1$M_{\odot}$ and mean [Fe/H] equal $-$0.35~dex, one with [$\alpha$/Fe] ([(Mg+Si)/Fe]) smaller than 0.07~dex (the scaled solar sample), 
the other one with average [$\alpha$/Fe]=0.20 (the $\alpha$-enhanced sample) respectively. 
These two sets of stars are distributed along well separated sequences, the $\alpha$-enhanced one being redder than the scaled solar sequence, as expected.
The $T_{\mathrm{eff}}$ difference between the two sequences is about 110~K at fixed log($g$).

We have also plotted 1.1$M_{\odot}$, [Fe/H]$-$0.35 models both scaled solar and with [$\alpha$/Fe]=0.4, from our own calculations and 
from \cite{dartmouth} isochrone database, for a comparison. For these latter models we have used the online 
webtool\footnote{http://stellar.dartmouth.edu/models/webtools.html} and calculated [Fe/H]=$-$0.35 isochrones populated by $\sim$1.1$_{\odot}$ 
stars along the RGB. 

The observed $T_{\mathrm{eff}}$ difference between scaled solar and $\alpha$-enhanced stars turns out to be reproduced by both independent sets of 
stellar models for [$\alpha$/Fe]$\sim$0.4, twice the observed value. 
This further analysis confirms that the trend of $\Delta T$ with [Fe/H] obtained with our models is due 
to the fact that they are systematically hotter than observations for $\alpha$-enhanced stars. Also, this discrepancy 
between $\alpha$-enhanced RGB stellar models and observations seems to be more general, not just related to our models.

\subsection{The effect of the solar metal distribution}\label{metdistr}

To assess better the good agreement between our scaled solar models and RGB sample ([$\alpha$/Fe]$<$0.07), we have also calculated a set of models 
with the same physics inputs but a more recent determination of the solar metal distribution (both opacities and 
equation of state take into account the new metal mixture), 
from \cite{caffau} for the most abundant elements, complemented with abundances from \cite{lodders}. We have covered the same range of masses and [Fe/H] of 
the reference models employed in the analysis described in the previous sections.
Notice that T17 calculations use the \cite{gs98} solar metal distribution, very similar to the \cite{gn93} one of our reference calculations.
The \cite{caffau} solar metal mixture implies a lower metallicity for the Sun compared to \cite{gn93}\footnote{The solar metallicity 
from the \cite{caffau} determination is slightly higher than what would be obtained with the \cite{asplund} solar metal mixture}. Our calibrated 
solar model provides an initial He abundance $Y$=0.269, metallicity $Z$=0.0172, $\alpha_{\rm MLT}$=2.0, and $\Delta Y$/$\delta Z$=1.31.

   \begin{figure}
   \centering
   \includegraphics[width=8.7cm]{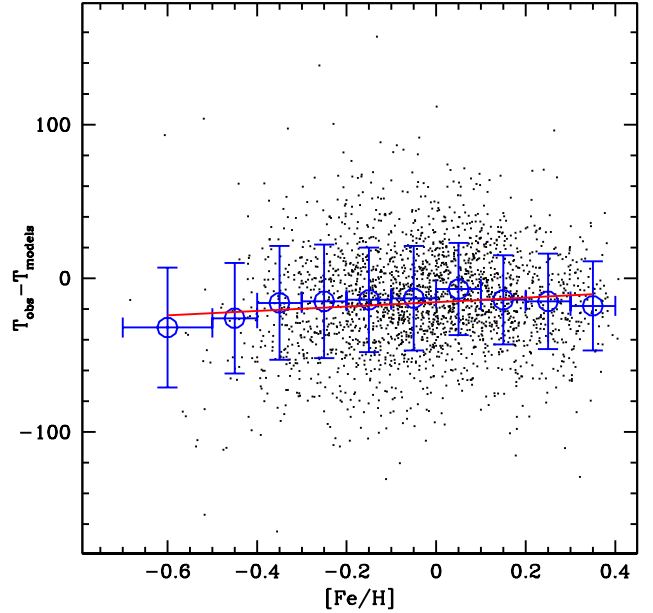}
      \caption{$\Delta T$ as a function of [Fe/H] for the sub-sample of RGB objects with 
      ${\rm [\alpha/Fe]<0.07}$, using stellar models calculated with the \cite{caffau} plus \cite{lodders} solar metal distribution (see text for details).}
         \label{dtcaffau}
   \end{figure}

Figure ~\ref{dtcaffau} displays the $\Delta T$ values as a function of [Fe/H] for the scaled solar sample. 
We have binned the $\Delta T$ values for [Fe/H] larger than $\sim -$0.6~dex (a [Fe/H] range of about 1~dex), and 
performed a linear fit to the binned data as in Fig.~\ref{differences}, deriving 
a slope equal to 14 $\pm$ 10 K/dex, statistically different from zero at much less than 2$\sigma$. 
The average $\Delta T$ is equal to $-$14~K, with a 1$\sigma$ dispersion of 34~K. We can conclude that changing the reference solar metal distribution 
and the corresponding solar calibrated $\alpha_{\rm MLT}$ does not 
alter the agreement between our models and the $T_{\mathrm{eff}}$ of the scaled solar T17 sample of RGB stars.

\subsection{The effect of the model boundary conditions}\label{bcon}

As discussed in \cite{kww}, the outer boundary conditions for the solution of the stellar evolution equations have a major effect 
on models with deep convective envelopes, like the RGB ones. We have therefore explored in some detail this issue, to check whether different choices 
of how the boundary conditions are determined can 
cause metallicity dependent $T_{\mathrm{eff}}$ differences amongst models with the same total mass, and eventually --at least partially-- explain the differences between our 
and T17 results.

The physics inputs of BaSTI and T17 calculations are very similar, 
the main difference being their integration of the Eddington grey $T(\tau)$ 
to determine the boundary conditions, whereas we used the VAL $T(\tau)$\footnote{The equation 
of state (EOS) is also different \citep[see T17 and][]{basti}, but tests made by \citet{basti} have shown 
that the EOS employed by T17 produces tracks very close to the ones obtained with the BaSTI EOS choice}.
We have therefore investigated the role played by different $T(\tau)$ choices to determine the outer boundary conditions of 
our model calculations \citep[see also, e.g.][for investigations of the effect of boundary conditions on the $T_{\mathrm{eff}}$ of 
low-mass stellar models with convective envelopes]
{montalban04, vdb08}, when comparing theory with the measured $T_{\mathrm{eff}}$ of T17 sample.

   \begin{figure}
   \centering
   \includegraphics[width=8.7cm]{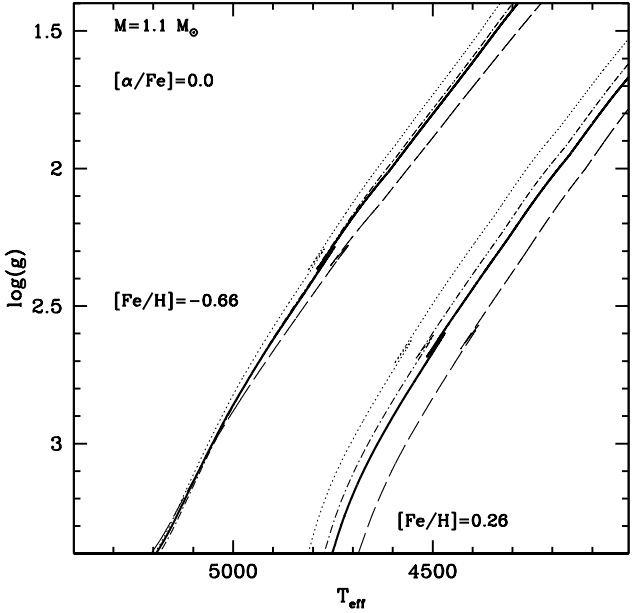}
      \caption{Scaled solar evolutionary tracks for the labelled mass and [Fe/H] values.  
     The tracks correspond to models calculated with the KS (dotted lines), VAL (solid lines), HM (dash-dotted lines) 
     and EDD (dashed lines) 
     $T(\tau)$ relationships, respectively (see text for details).}
         \label{boundary}
   \end{figure}

Figure~\ref{boundary} displays two groups of four RGB tracks in the 
$T_{\mathrm{eff}}$-log($g$) diagram (within the $T_{\mathrm{eff}}$ and log($g$) range sampled by T17 data), 
for 1.1$M_{\odot}$ models with the labelled [Fe/H] (scaled solar 
metal mixtures). The two chosen [Fe/H] values bracket 
the metallicity range covered by our $\Delta T$-[Fe/H] analysis, and 
the four tracks for each [Fe/H] represent four different choices for the $T(\tau)$ relation used to determine the outer boundary conditions.

Our reference calculations employing the VAL $T(\tau)$ 
are plotted together with calculations using an Eddington grey $T(\tau)$ (hereafter EDD) like T17 models, the \citet{ks} 
(hereafter KS) and the \citet{hm74} (hereafter HM) one\footnote{We employed 
the analytical fit by \citet{vdbp89} to the \citet{hm74} data}. 
The KS and HM $T(\tau)$ relationships are also solar semi-empirical, like the VAL $T(\tau)$. 

Values of $\alpha_{\rm MLT}$ for these additional models 
have been fixed again by means of a solar calibration, and are equal to 1.70  
(very close to the value 1.72 determined by T17 with their own calculations) 2.11 and 1.99 
for calculations with the EDD, KS and HM $T(\tau)$, respectively. For the sake of comparison, we remind the reader that the 
solar calibration with the VAL $T(\tau)$ requires $\alpha_{\rm MLT}=1.90$/

It is striking not only that different $T(\tau)$ relations and their corresponding solar calibrated $\alpha_{\rm MLT}$ values produce 
RGBs with different $T_{\mathrm{eff}}$ \citep[this was already shown for example in][]{scw02},  
but also that differences depend on the model [Fe/H]. These results are qualitatively and quantitatively the same also 
for masses equal to 2.0-2.5 $M_{\odot}$, at the upper end of the mass range spanned by T17 data.
Clearly, different solar calibrations of $\alpha_{\rm MLT}$ obtained with different $T(\tau)$ relations do not guarantee consistent behaviours  
of RGB models with metallicity.

At [Fe/H]=+0.26, all tracks are roughly parallel. The EDD track is cooler by $\sim$70~K compared to the reference VAL one, whereas the KS track is hotter 
by about the same amount, and the HM track is hotter by just $\sim$25~K. 
At solar metallicity (not shown in the figure) the differences between EDD, VAL, KS and HM tracks are still about the same as at [Fe/H]=+0.26, whilst 
at [Fe/H]=$-$0.66 EDD, VAL and KS tracks are no longer parallel. Above log($g$)$\sim$2.8 they are roughly coincident, with $T_{\mathrm{eff}}$ differences 
increasing with decreasing log($g$). At log($g$)=1.5 the EDD track is cooler by $\sim$40~K, while the KS track is hotter by $\sim$30~K. 
The HM track is almost coincident with the VAL one.

We have seen before that $\Delta T$ values determined from our calculations employing the VAL $T(\tau)$ relation do not 
show any trend with $T_{\mathrm{eff}}$, and are consistent with zero when scaled solar stars are considered. 
Employing instead the EDD $T(\tau)$ would increase $\Delta T$ values by $\sim$70~K at the upper end of the sampled [Fe/H] range down to about solar 
(EDD tracks being 
systematically cooler than VAL tracks), whilst the increase ranges from negligible to at most 40~K (when gravity decreases) 
at the lowest end of the [Fe/H] range considered in our analysis.
This would induce an overall positive trend in the $\Delta T$-[Fe/H] diagram ( $\Delta T$  
decreasing with decreasing [Fe/H]) also when restricting the analysis to scaled solar objects, 
with absolute $\Delta T$ values generally positive. This is at least qualitatively consistent with T17 results, even though  
it does not fully explain quantitatively T17 results, especially the very large positive $\Delta T$ at solar metallicity. Notice that also the PARSEC 
calculations --that according to T17 study show also a $\Delta T$-[Fe/H] slope of about 100~K/dex--  
employ an Eddington grey $T(\tau)$ relationship to determine the model outer boundary conditions. 

Finally, we is also interesting to notice that \cite{dartmouth} models display $T_{\mathrm{eff}}$ values very close to ours over the whole mass, surface gravity and 
[Fe/H] range covered by our analysis (see also Fig.~\ref{dtaenfe}). In those models the boundary conditions have been taken 
from a grid of PHOENIX detailed 1D model atmospheres \citep[pressure and temperature at a given optical depth $\tau$, see][]{dartmouth} instead of 
a $T(\tau)$ integration.

\section{Summary and discussion} \label{s:discussion}

Our reanalysis of the T17 sample of RGB stars from the APOKASC catalogue has disclosed the following:

\begin{enumerate}
\item{According to the APOKASC $T_{\mathrm{eff}}$, log($g$), mass, [Fe/H] and [$\alpha$/Fe] values given by T17, 
theoretical stellar evolution calculations --both our own calculations and T17 models, and also \cite{dartmouth} calculations-- seem to  
underestimate the effect of $\alpha$-enhancement on the model $T_{\mathrm{eff}}$ at fixed mass, surface gravity and [Fe/H].}
\item{When $\alpha$-enhanced stars are neglected, our RGB models 
are in good agreement with the empirical $T_{\mathrm{eff}}$ values, with no significant systematic shifts, nor trends with [Fe/H], 
over a $\sim$1~dex [Fe/H] range ([Fe/H] between $\sim+$0.4 and $\sim-$0.6).  
This agreement is preserved also if we change the reference solar metal distribution of our models.} 
\item{For a solar calibrated $\alpha_{\rm MLT}$, the $T_{\mathrm{eff}}$ differences between theory and observations 
depend on the choice of the model boundary conditions. It is the combinations of boundary conditions 
and $\alpha_{\rm MLT}$ value that determine the $T_{\mathrm{eff}}$ of RGB stellar models, as expected for stars with deep convective envelopes 
\citep{kww}.}
\end{enumerate}

Regarding the discrepancy between our models and $\alpha$-enhanced stars, a variation of  $\alpha_{\rm MLT}$ with [$\alpha$/Fe] 
at a given [Fe/H] seems unlikely --but of course cannot be a-priori dismissed. Another possibility that we have checked is the role played 
by the $\Delta Y/ \Delta Z$ enrichment ratio used in the model calculations. This value is typically fixed by the assumed 
primordial He and the solar initial $Y$ (and $Z$) obtained from a standard solar model. For a fixed value of [Fe/H], $\alpha$-enhanced stars have a larger $Z$, 
hence the corresponding models will have been calculated with a larger $Y$ compared to the scaled solar counterparts at the same [Fe/H] 
\citep[see, e.g. Table~3 in][]{dartmouth}. What if the initial $Y$ of $\alpha$-enhanced stars is the same as for the scaled solar ones at a given [Fe/H]? 
In the [Fe/H] range of T17 stars and for the observed $\alpha$-enhancements, the initial $Y$ of the 
$\alpha$-enhanced models will be at most $\sim$0.01 larger at the same [Fe/H], according to the $\Delta Y/ \Delta Z$ value used in 
our calculations. This small variation of $Y$ increases the $T_{\mathrm{eff}}$ of the models by $\sim$20~K in the relevant $g$ range, hence our 
$\alpha$-enhanced models would be at most 20~K cooler at fixed [Fe/H], if $Y$ is the same as for the scaled solar stars. 
Such a small change of the model $T_{\mathrm{eff}}$ would not erase the discrepancy with observations. 
  
From the empirical point of view, assuming metal abundance, $g$ and $T_{\mathrm{eff}}$ scales are correct, 
asteroseismic masses systematically too high by 0.1-0.2~$M_{\odot}$ for $\alpha$-enhanced stars could explain the discrepancy. 
Another possibility --  assuming mass, $g$ and $T_{\mathrm{eff}}$ values are correct-- 
is that [Fe/H] determinations are too low by $\sim$0.1~dex for each 0.1~dex of $\alpha$-enhancement, 
or a combination of both mass and [Fe/H] systematic errors.  
Of course it is necessary also to investigate this problem with stellar evolution models, to see whether there is room to 
explain this discrepancy from the theoretical side.

Concerning possible mixing length variations with chemical composition, we conclude that to match T17 $T_{\mathrm{eff}}$ values for the scaled solar sub-sample, 
variations of $\alpha_{\rm MLT}$ with [Fe/H] are required only for some choices of the model outer boundary conditions. 
Depending on the chosen $T(\tau)$ relation, or more in general the chosen set of boundary conditions, 
a variation of $\alpha_{\rm MLT}$ with [Fe/H] may or may not be necessary. Trying to determine whether $\alpha_{\rm MLT}$ 
can be assumed constant irrespective of chemical composition (and mass) for RGB stars thus requires a definitive assessment of 
the most correct way to determine the model outer boundary conditions.

As a consequence, models calculated with 
$\alpha_{\rm MLT}$ calibrations based on 3D radiation hydrodynamics simulations \citep{trampedach14, mwa15} 
are physically consistent only when boundary conditions (and physics inputs) extracted from the same simulations are employed in the 
stellar model calculations.
We have achieved this consistency in \citet{sc15}, testing the impact of \citet{trampedach14} simulations on stellar modelling.

\citet{sc15} have shown that at solar metallicity --the single metallicity covered by these 3D calculations-- 
RGB models calculated with the hydro-calibrated variable 
$\alpha_{\rm MLT}$ 
(that is a function of $T_{\mathrm{eff}}$ and log($g$)) are consistent --within about 20~K-- with RGB tracks obtained with the solar $\alpha_{\rm MLT}$ derived from the 
same set of hydro-simulations.
They also found that the VAL $T(\tau)$ relationship provides RGB effective temperatures 
that agree quite well with results obtained with the hydro-calibrated $T(\tau)$ relationship, within typically 10~K.
Assuming these hydro-simulations are realistic and accurate, the use of the VAL $T(\tau)$ and solar calibrated $\alpha_{\rm MLT}$ seems to be adequate 
for RGB stars at solar [Fe/H].

The independent 3D hydro-simulations by \citet{mwa15} cover a large [Fe/H] range, from $-$4.0 to +0.5, and provide corresponding calibrations 
of $\alpha_{\rm MLT}$ in terms of [Fe/H], $T_{\mathrm{eff}}$, and log($g$). 
However, $T(\tau)$ relationships (or tables of boundary conditions) obtained from their simulations, plus  
Rosseland mean opacities consistent with the opacities used in their calculations, are not yet available, 
This means that their $\alpha_{\rm MLT}$ calibration cannot be consistently implemented 
in stellar evolution calculations yet, and one cannot yet check consistently whether the variable $\alpha_{\rm MLT}$ 
provides RGB $T_{\mathrm{eff}}$ values significantly different from the case a solar hydro-calibrated $\alpha_{\rm MLT}$ for the full range of [Fe/H] covered by 
these simulations.

\begin{acknowledgements}
SC acknowledges financial support from PRIN-INAF2014 (PI: S. Cassisi) and the Economy and Competitiveness Ministry of the Kingdom of Spain (grant AYA2013-42781-P).
\end{acknowledgements}

\bibliographystyle{aa}

\begin{thebibliography}{34}
\expandafter\ifx\csname natexlab\endcsname\relax\def\natexlab#1{#1}\fi

\bibitem[{{Asplund} {et~al.}(2009){Asplund}, {Grevesse}, {Sauval}, \&
  {Scott}}]{asplund}
{Asplund}, M., {Grevesse}, N., {Sauval}, A.~J., \& {Scott}, P. 2009, \araa, 47,
  481

\bibitem[{{B{\"o}hm-Vitense}(1958)}]{bv58}
{B{\"o}hm-Vitense}, E. 1958, \zap, 46, 108

\bibitem[{{Bressan} {et~al.}(2012){Bressan}, {Marigo}, {Girardi}, {Salasnich},
  {Dal Cero}, {Rubele}, \& {Nanni}}]{parsec}
{Bressan}, A., {Marigo}, P., {Girardi}, L., {et~al.} 2012, \mnras, 427, 127

\bibitem[{{Caffau} {et~al.}(2011){Caffau}, {Ludwig}, {Steffen}, {Freytag}, \&
  {Bonifacio}}]{caffau}
{Caffau}, E., {Ludwig}, H.-G., {Steffen}, M., {Freytag}, B., \& {Bonifacio}, P.
  2011, \solphys, 268, 255

\bibitem[{{Dotter} {et~al.}(2008){Dotter}, {Chaboyer}, {Jevremovi{\'c}},
  {Kostov}, {Baron}, \& {Ferguson}}]{dartmouth}
{Dotter}, A., {Chaboyer}, B., {Jevremovi{\'c}}, D., {et~al.} 2008, \apjs, 178,
  89

\bibitem[{{Ferguson} {et~al.}(2005){Ferguson}, {Alexander}, {Allard}, {Barman},
  {Bodnarik}, {Hauschildt}, {Heffner-Wong}, \& {Tamanai}}]{ferguson}
{Ferguson}, J.~W., {Alexander}, D.~R., {Allard}, F., {et~al.} 2005, \apj, 623,
  585

\bibitem[{{Gonz{\'a}lez Hern{\'a}ndez} \& {Bonifacio}(2009)}]{ghb09}
{Gonz{\'a}lez Hern{\'a}ndez}, J.~I. \& {Bonifacio}, P. 2009, \aap, 497, 497

\bibitem[{{Gough} \& {Weiss}(1976)}]{gw76}
{Gough}, D.~O. \& {Weiss}, N.~O. 1976, \mnras, 176, 589

\bibitem[{{Grevesse} \& {Noels}(1993)}]{gn93}
{Grevesse}, N. \& {Noels}, A. 1993, Physica Scripta Volume T, 47, 133

\bibitem[{{Grevesse} \& {Sauval}(1998)}]{gs98}
{Grevesse}, N. \& {Sauval}, A.~J. 1998, \ssr, 85, 161

\bibitem[{{Holweger} \& {Mueller}(1974)}]{hm74}
{Holweger}, H. \& {Mueller}, E.~A. 1974, \solphys, 39, 19

\bibitem[{{Iglesias} \& {Rogers}(1996)}]{ir:96}
{Iglesias}, C.~A. \& {Rogers}, F.~J. 1996, \apj, 464, 943

\bibitem[{{Kippenhahn} {et~al.}(2012){Kippenhahn}, {Weigert}, \& {Weiss}}]{kww}
{Kippenhahn}, R., {Weigert}, A., \& {Weiss}, A. 2012, {Stellar Structure and
  Evolution}

\bibitem[{{Krishna Swamy}(1966)}]{ks}
{Krishna Swamy}, K.~S. 1966, \apj, 145, 174

\bibitem[{{Lodders}(2010)}]{lodders}
{Lodders}, K. 2010, Astrophysics and Space Science Proceedings, 16, 379

\bibitem[{{Ludwig} {et~al.}(1999){Ludwig}, {Freytag}, \& {Steffen}}]{lfs}
{Ludwig}, H.-G., {Freytag}, B., \& {Steffen}, M. 1999, \aap, 346, 111

\bibitem[{{Magic} {et~al.}(2015){Magic}, {Weiss}, \& {Asplund}}]{mwa15}
{Magic}, Z., {Weiss}, A., \& {Asplund}, M. 2015, \aap, 573, A89

\bibitem[{{Majewski} {et~al.}(2017){Majewski}, {Schiavon}, {Frinchaboy},
  {Allende Prieto}, {Barkhouser}, {Bizyaev}, {Blank}, {Brunner}, {Burton},
  {Carrera}, {Chojnowski}, {Cunha}, {Epstein}, {Fitzgerald}, {Garc{\'{\i}}a
  P{\'e}rez}, {Hearty}, {Henderson}, {Holtzman}, {Johnson}, {Lam}, {Lawler},
  {Maseman}, {M{\'e}sz{\'a}ros}, {Nelson}, {Nguyen}, {Nidever}, {Pinsonneault},
  {Shetrone}, {Smee}, {Smith}, {Stolberg}, {Skrutskie}, {Walker}, {Wilson},
  {Zasowski}, {Anders}, {Basu}, {Beland}, {Blanton}, {Bovy}, {Brownstein},
  {Carlberg}, {Chaplin}, {Chiappini}, {Eisenstein}, {Elsworth}, {Feuillet},
  {Fleming}, {Galbraith-Frew}, {Garc{\'{\i}}a}, {Garc{\'{\i}}a-Hern{\'a}ndez},
  {Gillespie}, {Girardi}, {Gunn}, {Hasselquist}, {Hayden}, {Hekker}, {Ivans},
  {Kinemuchi}, {Klaene}, {Mahadevan}, {Mathur}, {Mosser}, {Muna}, {Munn},
  {Nichol}, {O'Connell}, {Parejko}, {Robin}, {Rocha-Pinto}, {Schultheis},
  {Serenelli}, {Shane}, {Silva Aguirre}, {Sobeck}, {Thompson}, {Troup},
  {Weinberg}, \& {Zamora}}]{maj}
{Majewski}, S.~R., {Schiavon}, R.~P., {Frinchaboy}, P.~M., {et~al.} 2017, \aj,
  154, 94

\bibitem[{{Montalb{\'a}n} {et~al.}(2004){Montalb{\'a}n}, {D'Antona}, {Kupka},
  \& {Heiter}}]{montalban04}
{Montalb{\'a}n}, J., {D'Antona}, F., {Kupka}, F., \& {Heiter}, U. 2004, \aap,
  416, 1081

\bibitem[{{Pedersen} {et~al.}(1990){Pedersen}, {Vandenberg}, \&
  {Irwin}}]{pvi90}
{Pedersen}, B.~B., {Vandenberg}, D.~A., \& {Irwin}, A.~W. 1990, \apj, 352, 279

\bibitem[{{Pietrinferni} {et~al.}(2004){Pietrinferni}, {Cassisi}, {Salaris}, \&
  {Castelli}}]{basti}
{Pietrinferni}, A., {Cassisi}, S., {Salaris}, M., \& {Castelli}, F. 2004, \apj,
  612, 168

\bibitem[{{Salaris} \& {Cassisi}(1996)}]{sc:96}
{Salaris}, M. \& {Cassisi}, S. 1996, \aap, 305, 858

\bibitem[{{Salaris} \& {Cassisi}(2008)}]{sc08}
{Salaris}, M. \& {Cassisi}, S. 2008, \aap, 487, 1075

\bibitem[{{Salaris} \& {Cassisi}(2015)}]{sc15}
{Salaris}, M. \& {Cassisi}, S. 2015, \aap, 577, A60

\bibitem[{{Salaris} {et~al.}(2002){Salaris}, {Cassisi}, \& {Weiss}}]{scw02}
{Salaris}, M., {Cassisi}, S., \& {Weiss}, A. 2002, \pasp, 114, 375

\bibitem[{{Stein} \& {Nordlund}(1989)}]{sn}
{Stein}, R.~F. \& {Nordlund}, A. 1989, \apjl, 342, L95

\bibitem[{{Straniero} \& {Chieffi}(1991)}]{sc:91}
{Straniero}, O. \& {Chieffi}, A. 1991, \apjs, 76, 525

\bibitem[{{Tayar} {et~al.}(2017){Tayar}, {Somers}, {Pinsonneault}, {Stello},
  {Mints}, {Johnson}, {Zamora}, {Garc{\'{\i}}a-Hern{\'a}ndez}, {Maraston},
  {Serenelli}, {Allende Prieto}, {Bastien}, {Basu}, {Bird}, {Cohen}, {Cunha},
  {Elsworth}, {Garc{\'{\i}}a}, {Girardi}, {Hekker}, {Holtzman}, {Huber},
  {Mathur}, {M{\'e}sz{\'a}ros}, {Mosser}, {Shetrone}, {Silva Aguirre},
  {Stassun}, {Stringfellow}, {Zasowski}, \& {Roman-Lopes}}]{tayar}
{Tayar}, J., {Somers}, G., {Pinsonneault}, M.~H., {et~al.} 2017, \apj, 840, 17

\bibitem[{{Trampedach} {et~al.}(2014){Trampedach}, {Stein},
  {Christensen-Dalsgaard}, {Nordlund}, \& {Asplund}}]{trampedach14}
{Trampedach}, R., {Stein}, R.~F., {Christensen-Dalsgaard}, J., {Nordlund},
  {\AA}., \& {Asplund}, M. 2014, \mnras, 445, 4366

\bibitem[{{VandenBerg} {et~al.}(2012){VandenBerg}, {Bergbusch}, {Dotter},
  {Ferguson}, {Michaud}, {Richer}, \& {Proffitt}}]{vdb12}
{VandenBerg}, D.~A., {Bergbusch}, P.~A., {Dotter}, A., {et~al.} 2012, \apj,
  755, 15

\bibitem[{{Vandenberg} {et~al.}(1996){Vandenberg}, {Bolte}, \&
  {Stetson}}]{vbs96}
{Vandenberg}, D.~A., {Bolte}, M., \& {Stetson}, P.~B. 1996, \araa, 34, 461

\bibitem[{{VandenBerg} {et~al.}(2008){VandenBerg}, {Edvardsson}, {Eriksson}, \&
  {Gustafsson}}]{vdb08}
{VandenBerg}, D.~A., {Edvardsson}, B., {Eriksson}, K., \& {Gustafsson}, B.
  2008, \apj, 675, 746

\bibitem[{{Vandenberg} \& {Poll}(1989)}]{vdbp89}
{Vandenberg}, D.~A. \& {Poll}, H.~E. 1989, \aj, 98, 1451

\bibitem[{{Vernazza} {et~al.}(1981){Vernazza}, {Avrett}, \&
  {Loeser}}]{vernazza:81}
{Vernazza}, J.~E., {Avrett}, E.~H., \& {Loeser}, R. 1981, \apjs, 45, 635

\end{thebibliography}

\end{document}